\documentstyle[11pt,paspconf,epsf]{article}
\def\lsim{\lower.5ex\hbox{$\; \buildrel < \over \sim \;$}}
\def\gsim{\lower.5ex\hbox{$\; \buildrel > \over \sim \;$}}

\def\edcomment#1{\iffalse\marginpar{\raggedright\sl#1\/}\else\relax\fi}
\reversemarginpar

\begin{document}
\title{Spallogenic Light Elements and Cosmic-Ray Origin}

\author{Reuven Ramaty}
\affil{NASA/GSFC, Greenbelt, MD 20771}

\author{Richard E. Lingenfelter}
\affil{CASS/UCSD, LaJolla, CA 92093}

\begin{abstract}

Using a Monte Carlo code which incorporates hitherto ignored
effects, the delayed mixing into the ISM of the nucleosynthetic
yields of supernovae and the relationship between these yields and
the supernova ejecta kinetic energy (including events with very
large ejecta energies, e.g. collapsars/hypernovae), we tested the
viability of the three recent evolutionary models for the origin of
the spallogenic light elements, specifically Be. We find that the
CRI, in which the cosmic rays are accelerated out of an ISM which is
increasingly metal poor at early times, significantly under-predicts
the measured Be abundance at the lowest [Fe/H], the increase in
[O/Fe] with decreasing [Fe/H] indicated by recent data
notwithstanding. We also find that the CRS, in which cosmic rays
with composition similar to that of the current epoch cosmic rays
are accelerated out of fresh supernova ejecta and thus maintain a
constant metallicity at all epochs, can account for the measured Be
abundances with an acceleration efficiency that is in good agreement
with the current epoch cosmic-ray data. In addition, we show that
the LECR, in which a postulated low energy component also
accelerated out of fresh ejecta coexists with the CRI cosmic rays,
can account for the observations as well, except that the required
acceleration efficiency becomes prohibitively large for some of the
parameters that have been previously assumed for this as-yet
undetected cosmic-ray component.

\end{abstract}

\section{Introduction}

Recent measurements (see Vangioni-Flam et al. 1998, hereafter
VF98) of Be and B abundances in low-metallicity stars provide
important new clues as to the source of the cosmic-ray nuclei that
are in conflict with the current paradigm (hereafter the CRI
model) of cosmic-ray acceleration out of the average interstellar
medium (ISM). In particular, these measurements show that the
ratio of spallation-produced Be to supernova-nucleosynthetic Fe is
very nearly independent of metallicity, rather than being
proportional to the metallicity, as would be expected if all the
cosmic rays were accelerated out of the average ISM.

To explain these results, we proposed (Ramaty et al. 1997; Ramaty,
Kozlovsky \& Lingenfelter 1998a; Lingenfelter, Ramaty \& Kozlovsky
1998; Higdon, Lingenfelter \& Ramaty 1998; Ramaty, Lingenfelter \&
Kozlovsky 1998b) an alternative model for the origin of the cosmic
rays and of the spallogenic light elements, $^6$Li, Be and part of
the B. Specifically we developed a model (hereafter CRS) in which
the cosmic rays are accelerated out of fresh supernova ejecta and at
all epochs of Galactic evolution these light elements were produced
by spallation interactions of such cosmic rays with the ambient ISM.
This model differs from the traditional CRI approach in which the
present epoch cosmic rays are accelerated out an ambient medium of
solar composition (suggested to be the ISM, Meyer, Drury \& Ellison
1997), and at all past epochs the composition of the source
particles of the cosmic rays that produce the light elements (e.g.
Prantzos, Cass\'e, \& Vangioni-Flam 1993) was that of the average
ISM at that epoch. However, the new Be abundance data show that the
dependence of log(Be/H) on [Fe/H]$\equiv$log(Fe/H)-log(Fe/H)$_\odot$
is essentially linear, not quadratic as predicted by this model. As
an alternative, hybrid models for the origin of the spallogenic
light elements were developed (Cass\'e, Lehoucq \& Vangioni-Flam
1995; Vangioni-Flam et al. 1996; Ramaty, Kozlovsky, \& Lingenfelter
1996; Vangioni-Flam, Cass\'e, \& Ramaty 1997). In these models
(hereafter LECR), a separate low energy cosmic-ray component, also
accelerated out of fresh nucleosynthetic matter, was suggested to
coexists with the CRI cosmic rays and to dominate the Be and B
production, particularly in the early Galaxy. Both the CRS and LECR
models thus predict a linear evolution of the Be abundance, while
the CRI model does not. Fields \& Olive (1999a), however, recently
suggested that the latter model might still be viable, based on
their re-analysis of the data, including O in low metallicity stars
(Israelian, Garcia Lopez, \& Rebolo 1998).

The three models (CRI, CRS, LECR) involve different current epoch
cosmic-ray origin scenarios. While the CRI scenario posits
acceleration from the average ISM (Ellison, Drury \& Meyer 1997),
CRS assumes that the cosmic rays are accelerated out of fresh
supernova ejecta. We have shown (Lingenfelter et al. 1998) that the
standard arguments against such an ejecta origin (e.g. Meyer et al.
1997) can be answered (see \S 4), with the most likely scenario
involving the collective acceleration by successive supernova shocks
of ejecta-enriched matter in the interiors of superbubbles (Higdon
et al. 1998). This scenario can also account for the delay between
nucleosynthesis and acceleration (time scales of $\sim 10^5$ yr),
suggested by the recent $^{59}$Co and $^{59}$Ni observations (Binns
et al. 1999). Moreover, these hot, low density superbubbles are, in
fact, the ``hot phase" of the interstellar medium where shock
acceleration of cosmic rays is expected (e.g. Axford 1981; Bykov \&
Fleishman 1992) to be most effective because the energy losses of
the accelerated particles are minimized and the supernova shocks do
not suffer major radiative losses, as they would in a denser medium.

The LECR scenario was motivated by the reported (Bloemen et al.
1994; 1997), but recently retracted (Bloemen et al. 1999), detection
with COMPTEL/CGRO of C and O nuclear gamma-ray lines from the Orion
star formation region. These gamma rays were attributed to a
postulated low energy cosmic-ray component, which was highly
enriched in C and O relative to protons and $\alpha$ particles (see
Ramaty 1996 for review and Ramaty et al. 1996 for extensive
calculations of light element production by LECRs). It was suggested
(Bykov \& Bloemen 1994; Ramaty et al. 1996; Parizot, Cass\'e, \&
Vangioni-Flam 1997) that such enriched LECRs might be accelerated
out of metal-rich winds of massive stars and the ejecta of
supernovae from massive star progenitors which explode within the
bubble around the star formation region due to their very short
lifetimes. It was further suggested (Parizot et al. 1997) that
acceleration by an ensemble of shocks in superbubbles (Bykov \&
Fleishman 1992) would produce the distinct low energy cosmic-ray
component involved in the LECR model. But the cosmic-ray spectrum
predicted by Bykov \& Fleishman (1992) is identical to the Galactic
cosmic-ray spectrum except that it is cut-off (becomes harder) at
low energies. Following the reported detection of nuclear gamma-ray
lines from Orion, the Bykov \& Fleishman (1992) model was modified
(Bykov \& Bloemen 1994) by lowering the shock compression ratio, so
that the ensuing weak shocks would produce a much softer high energy
cosmic-ray spectrum which was necessary to explain the lack of the
pion-decay radiation that would have had to accompany the
now-withdrawn nuclear line emission. Thus, new gamma-ray line data
are needed to determine the role of the LECR component.

Concerning B, it is well known (e.g. Ramaty et al. 1997) that the
CRS and CRI models cannot account for the measured meteoritic
$^{11}$B/$^{10}$B. A likely solution is some additional $^{11}$B
production by $\nu$-induced $^{12}$C spallation in supernovae
(Woosley \& Weaver 1995). As we have dealt with such B production in
our previous studies, we do not consider B in the present paper.
Both the CRS and CRI models can account for the $^6$Li-to-Be
abundance ratio in meteorites. There are, however, indications that
at low metallicities $^6$Li/Be exceeds the meteoritic value (Smith,
Lambert, \& Nissen 1998). The origin of this excess, not discussed
in the present paper, remains uncertain (see, however, Vangioni-Flam
et al. 1999 and Fields \& Olive 1999b).

\begin{figure}
\plottwo{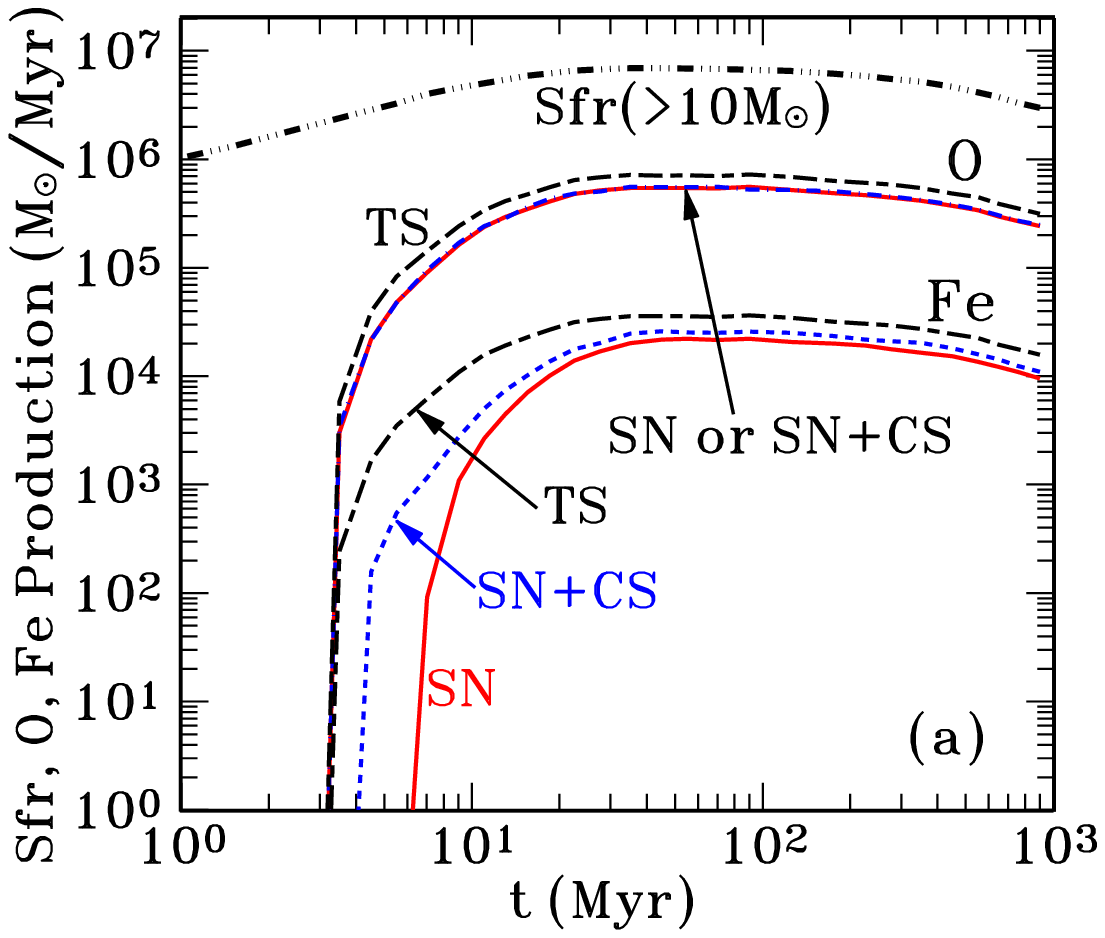}{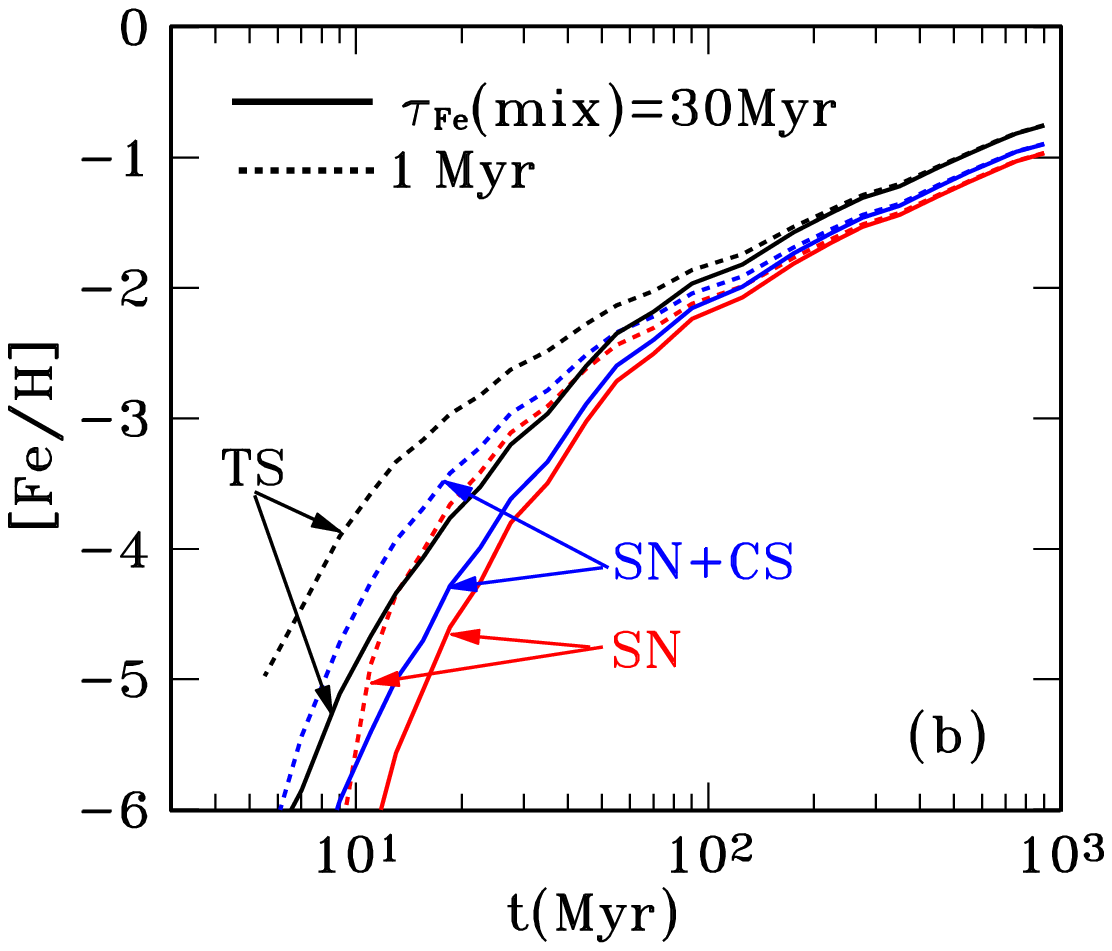}
\caption{Evolution of (a) - star ($>$10M$_\odot$) formation rate,
and O and Fe production rates; (b) - ISM Fe abundance for two Fe
mixing times with models (TS, SN and SN+CS) of Ni-Fe ejecta as
functions of core-collapse supernova progenitor mass (see text).}
\end{figure}

\section{Fe and O Production and Evolution}

We consider a one-zone model and limit our treatment to the first
10$^9$ years of Galactic evolution, because this low metallicity era
conveys the most unambiguous information on cosmic-ray origin. We
allow the ISM mass to accumulate exponentially, $d{\rm
M_{ISM}}(t)/dt = 10^{11}{\rm M_\odot} e^{-t/\tau}$. As we found that
the relevant abundance ratios are not significantly affected by the
value of $\tau$, we only show results for $\tau$=10 Myr.  We remove
ISM mass by star formation and outflow, taking the star formation
rate proportional to the ISM mass, $\Psi=\alpha{\rm M_{ISM}}$, and
the outflow proportional to $\Psi$, $O = \beta \Psi$ (Prantzos,
Cass\'e, \& Vangioni-Flam 1993). Again, since our results do not
depend appreciably on $\alpha$ and $\beta$ (see Ramaty et al.
1998b), we only show results for $\alpha = 0.6 {\rm Gyr}^{-1}$ and
$\beta = 1$. We set up a grid of 30
time bins, ranging from 0 to 1 Gyr, and accumulate mass into these
bins by the above prescription. Then for each bin we generate an
ensemble of stellar masses normalized to the assumed star formation
rate and distributed according to the Salpeter IMF up to 100
M$_\odot$. Starting from the first bin and advancing in time, we
calculate the removed mass and the returned mass for each star in
the ensemble, the latter depending on stellar lifetimes and the
remnant mass (e.g. Prantzos et al. 1993). We note that only
$>$2M$_\odot$ stars return mass for $t$$<$1 Gyr. The resultant star
formation rate for stars of mass exceeding 10M$_\odot$ is shown by
the top curve in Figure~1a.

\vskip -0.2truecm
\begin{table}
\caption{Progenitor and Ejected Oxygen and Iron ($^{56}$Ni) Masses
in M$_\odot$.}\label{tbl-1}
\begin{center}
\vskip -0.4truecm \scriptsize
\begin{tabular}{cccccccccccc}
M(progenitor)&12&13&15&18&20&22&25&30&35&40&$>$40\\ \tableline
M$_{\rm O}$(TS)\tablenotemark{a}
&0.065&0.12&0.4&0.7&1.3&1.8&2.9&4.5&6.0&8.0&12\\ M$_{\rm
Fe}$(TS)\tablenotemark{a}
&0.01&0.012&0.03&0.055&0.1&0.12&0.15&0.25&0.3&0.35&0.5\\ M$_{\rm
O}$(SN)\tablenotemark{b}
&0.15&0.18&0.38&0.77&1.51&1.73&2.78&4.07&5.55&6.2&6.2\\ M$_{\rm
Fe}$(SN)\tablenotemark{b}
&0.054&0.089&0.064&0.16&0.09&0.12&0.20&0&0&0&0\\
\end{tabular}
\end{center}
\vskip -0.85truecm \tablenotetext{a}{Shigeyama \& Tsujimoto 1998;
Tsujimoto \& Shigeyama 1998} \tablenotetext{b}{Woosley \& Weaver
1995}
\end{table}

Galactic Fe and O production is due to the $>$10 M$_\odot$ stars
which explode as core collapse supernovae. In each time bin and for
each star in the ensemble, we calculate the ejected O and Fe masses
for three cases. The first case (TS) is based on the yields given
by  Shigeyama \& Tsujimoto (1998) and Tsujimoto \& Shigeyama (1998),
which represent the best observationally determined values of
observed abundance patterns in extremely low metallicity stars.
These are bracketed by the yields of Woosley \& Weaver (1995,
hereafter WW95) at metallicity 10$^{-4}$ of the solar value for high
($>$20 M$_\odot$) mass progenitors with ejecta kinetic energy,
W$_{\rm SN}$, ranging from about 1 to 3$\times$10$^{51}$ erg. In the
second case (SN), which we chose in order to maximize the relative
Be production efficiency, we take the minimum Ni-Fe yield models of
WW95, which practically vanish at progenitor masses of 30 M$_\odot$
and above. Specifically, we use the ejecta corresponding to the
lowest W$_{\rm SN}$ at 30 M$_\odot$, and those at intermediate
W$_{\rm SN}$ at 35 and 40 M$_\odot$, because these still yield
vanishing Ni-Fe yields but finite O yields, thereby providing a good
fit to the O-to-Fe abundance ratio, as we shall see. As no
calculations for progenitors more massive than 40 M$_\odot$ are
available in WW95, we assume that the ejected O and Fe masses for
such progenitors are equal to those of 40 M$_\odot$. We summarize
the masses for both the TS and SN cases in Table~1. Even though
these yields depend on the W$_{\rm SN}$ employed by WW95, they are
within 30\% of $\sim$1.5$\times$10$^{51}$ erg, so we assume this
value for the entire progenitor mass range from 10 to 100 M$_\odot$
in both the TS and SN cases.

We also explore another case (SN+CS), based of the possible
existence of a rare class of much more energetic supernovae that
could enhance the Be production efficiency by imparting an order
of magnitude more energy to cosmic rays with only modestly
increased Fe yields. Such ``collapsars" (CS, also referred to as
hypernovae), whose ejecta jets are driven by black hole accretion
of ``failed supernovae", were suggested (Woosley 1993; Woosley,
Eastman \& Schmidt 1998; MacFadyen \& Woosley 1998) as possible
sources of gamma-ray bursts. The discovery of the Type Ic
supernova SN1998bw (Galama et al. 1998) in the error box of the
gamma-ray burst GRB980425 seems to support such a connection.
Various models (Hoflich, Wheeler \& Wang 1998; Iwamoto et al.
1998; Woosley, Eastman \& Schmidt 1998), with ejecta kinetic
energies W$_{\rm SN}$ ranging from 2 to 28$\times$10$^{51}$ erg,
are quite consistent with the observations of this supernova. We
incorporate such Type Ic supernovae by replacing 3\% of the
supernovae for the SN case by collapsars, based on their estimated
Galactic rate of 10$^{-3}$ yr$^{-1}$ (Hansen 1999). For the
remaining 97\%, we keep W$_{\rm SN}$ = 1.5$\times$10$^{51}$ erg
with ejected masses as given in Table~1, but to maximize the Be
production of the additional Type Ic's, we adopt the largest
ejecta energy in the Woosley et al. (1998) calculations, W$_{\rm
SN}$=2.8$\times$10$^{52}$ erg, corresponding to ejected $^{56}$Ni
and $^{16}$O masses of 0.49 and 2.8 M$_\odot$ respectively, and a
progenitor mass of 40 M$_\odot$. Since the rate of gamma-ray
bursts is only $\sim$10$^{-7}$ yr$^{-1}$ per galaxy, if their
emission is not beamed, the Type Ic rate of 10$^{-3}$ yr$^{-1}$
implies that only a very small fraction of the Type Ic supernovae
produce gamma-ray bursts, or that their gamma ray emission is
beamed into a very small solid angle.

The resultant O and Fe production rates as functions of time are
also shown in Figure~1a for the TS, SN and SN+CS cases. The delay of
a few Myr of the onset of the O and Fe production relative to the
star formation rate, due to the lifetimes of the progenitors, can be
clearly seen. There is an additional delay between the onset of O
and Fe production in the SN case, because here the supernovae from
the most massive, short-lived, progenitors produce O but not Fe. The
incorporation of the collapsars (CS) shortens this delay, but its
effect on the overall Fe production is quite small. There is no
delay between the O and Fe production in the TS case.

\begin{figure}
\plottwo{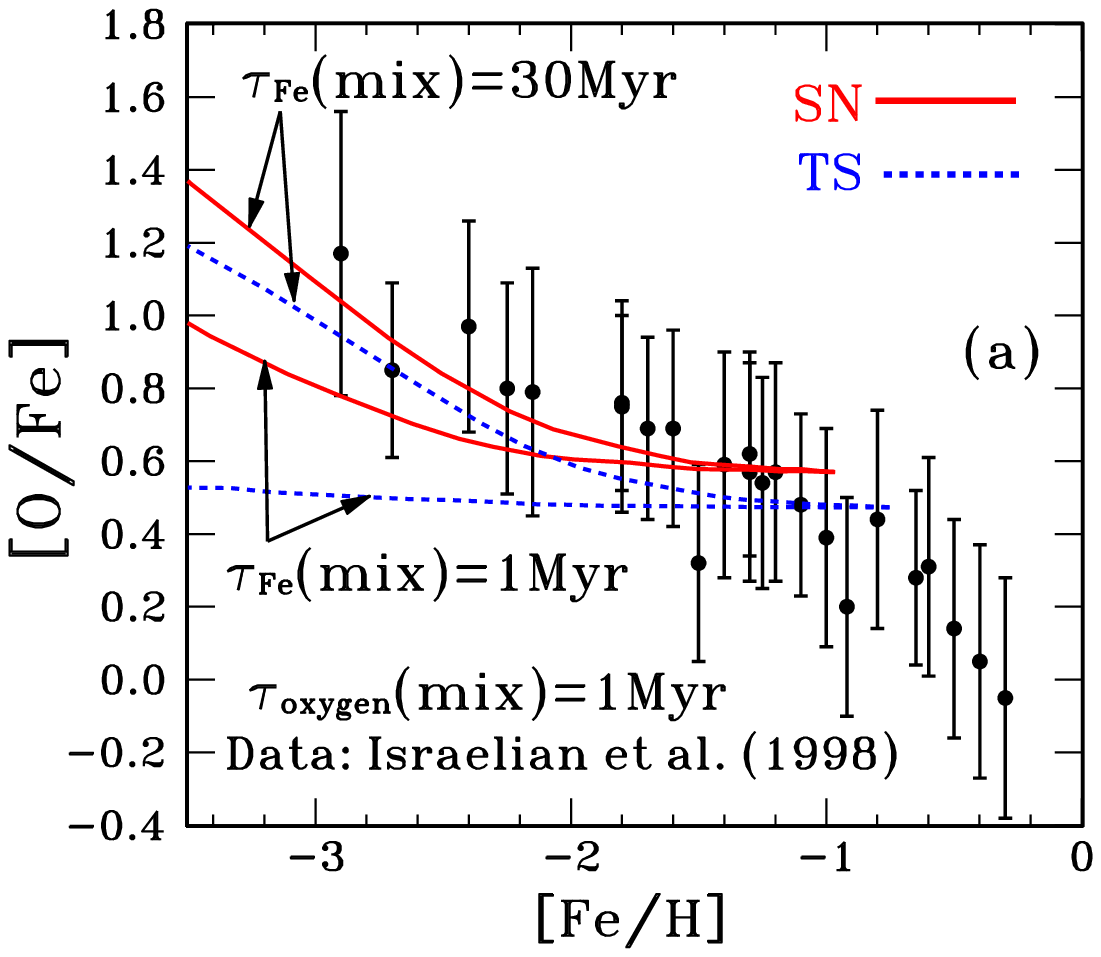}{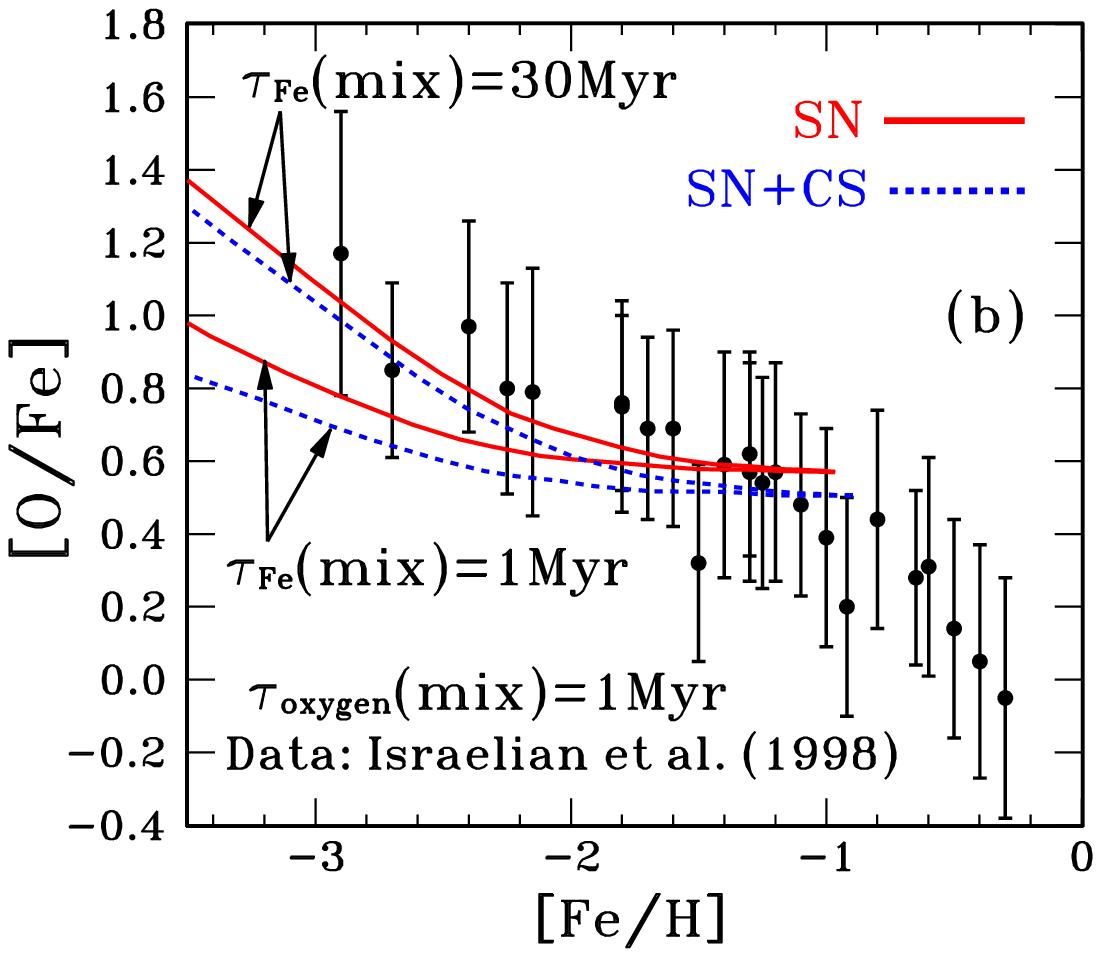}
\caption{O-to-Fe abundance ratio as a function of
[Fe/H]. TS, SN and SN+CS assume different O and Ni-Fe ejecta
yields as functions of core-collapse supernova progenitor mass; also
shown is the dependence of [O/Fe] on the Fe mixing time which could
become quite long as a result of the incorporation of the ejected Fe
into high velocity dust.}
\end{figure}

We calculate the O and Fe contents of the ISM by delaying, due to
transport and mixing, the deposition of these synthesized products.
In the simulation we choose the mixing times randomly in the
interval 0-$\tau({\rm mix})$ and allow for the possibility that
$\tau_{\rm Fe}$(mix)$>$$\tau_{\rm O}$(mix). We choose, arbitrarily,
a short mixing time for O, $\tau_{\rm O}$(mix)=1 Myr, but consider
two values for Fe, $\tau_{\rm Fe}$(mix)=1 and 30 Myr, the latter
based on a scenario in which the bulk of the ejected Fe would be
incorporated into high velocity refractory dust grains. This is
supported both by observations (Kozasa, Hasegawa \& Nomoto 1991)
suggesting the massive condensation of iron oxide and other
refractory grains at velocities $\simeq$2,500 km s$^{-1}$ in the
supernova 1987A and by observations (Naya et al. 1996) of the
Galactic 1.809 MeV gamma-ray line resulting from the decay of
$^{26}$Al, most likely produced in Type II supernovae (e.g. WW95).
The observed width of the Al line implies that the radioactive
aluminum is still moving at velocities $>$450 km s$^{-1}$ some
10$^6$ yrs after its formation and long after the associated
supernova remnants have slowed to thermal velocities, suggesting
that the bulk of the synthesized $^{26}$Al is in high velocity dust
grains which take a much longer time to slow down. Since both
aluminum and iron form highly refractory compounds, it is likely
that the bulk of the synthesized Fe also resides in such grains
which could travel much farther before they came to rest than the
volatiles trapped in the plasma which are more rapidly slowed by the
swept up gas and magnetic fields. Thus, there could be a delay
between the effective deposition times into the ISM of refractories,
such as Fe, and volatiles such as O, of which only a fraction is
condensed in oxide grains (see Lingenfelter et al. 1998 and \S 4).

The resulting evolution of the Fe abundance is shown in Figure~1b,
where we see that in 1 Gyr [Fe/H] indeed evolved to $\sim -1$. At
later epochs the contribution of thermonuclear supernovae (Type Ia)
becomes important (e.g. Kobayashi et al. 1998), but these are not
considered in the present paper. We see that [Fe/H] is largest for
the TS case, and at early times [Fe/H] for the SN case can be
increased by the CS augmentation. For all cases, as expected, [Fe/H]
is decreased by a longer Fe mixing time.

The calculated O-to-Fe abundance ratio
([O/Fe]$\equiv$log(O/Fe)-log(O/Fe)$_\odot$), together with recent
data (Israelian et al. 1998), are shown in Figure~2, where we
compare the predictions of the TS and SN cases (panel a) and those
of the SN and SN+CS cases (panel b). We see that if $\tau_{\rm
Fe}$(mix)=30 Myr, all cases are consistent with the data (the
decrease of [O/Fe] for [Fe/H]$> -1$ is due to the onset of
contributions from Type Ia supernovae which eject large Fe masses
but not much O). For smaller $\tau_{\rm Fe}$(mix), even though
[O/Fe] is lower at low [Fe/H], both the SN and SN+CS cases remain
consistent with the data and the inconsistency of the TS case is
limited to just the lowest values of [Fe/H]. It is interesting to
note that, unlike [O/Fe], the abundance ratios of the
$\alpha$-nuclei Mg, Si, Ca and Ti relative to Fe do not increase
with decreasing [Fe/H] below [Fe/H]=$-1$ (Ryan, Norris, \& Beers
1996). This may be consistent with the fact that these elements
are also refractory, and thus are affected by mixing in the same
way as is Fe.

\begin{figure}
\plottwo{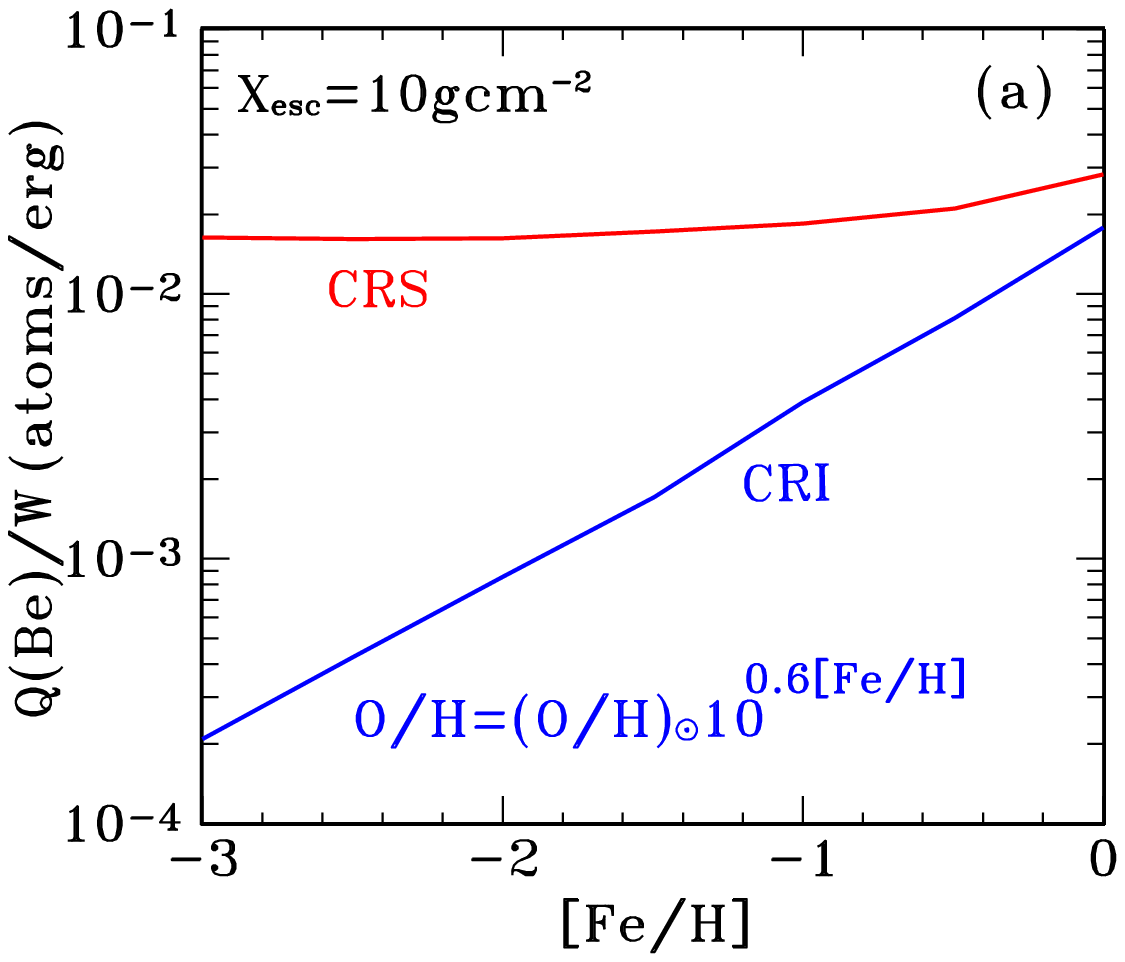}{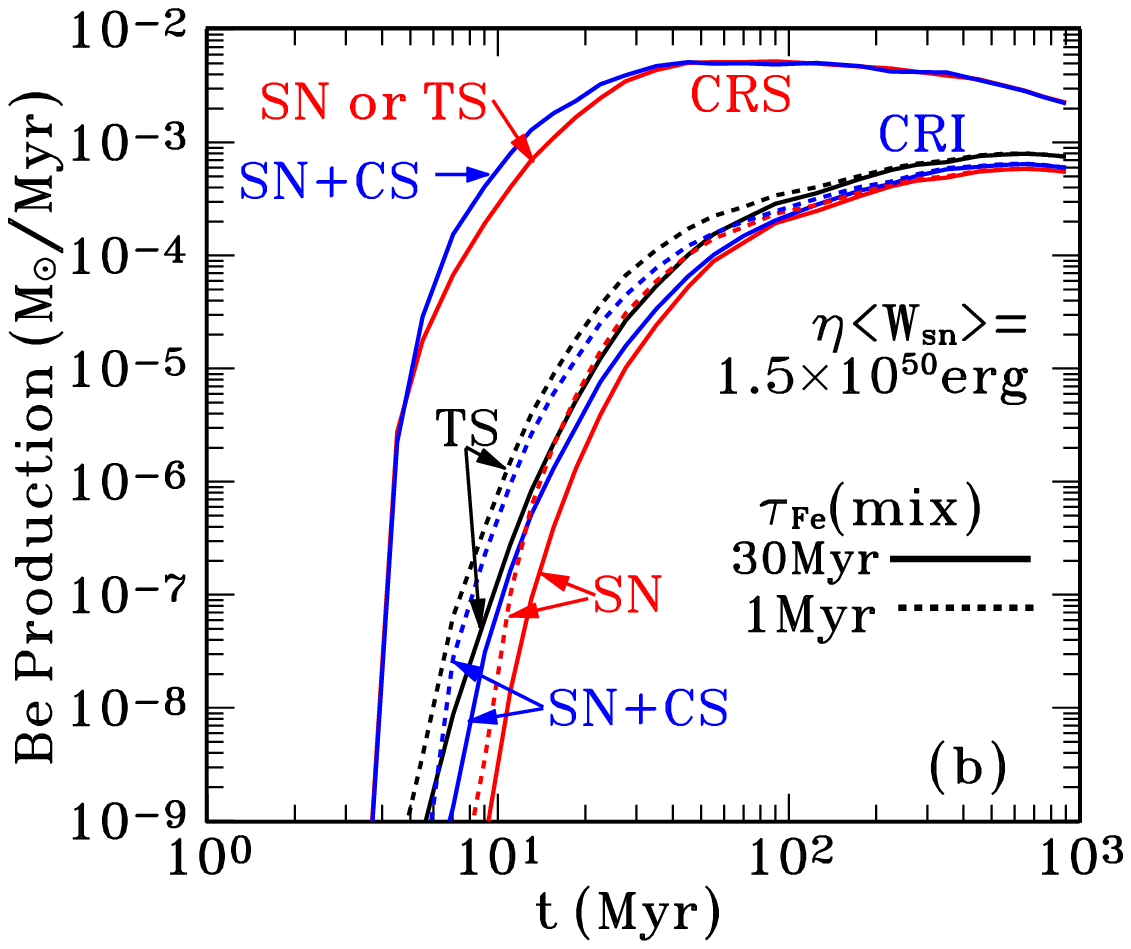}
\caption{(a) Number of Be atoms produced per unit
cosmic-ray energy for the CRS and CRI models; X$_{\rm esc}$ = 10 g
cm$^{-2}$ is the current epoch cosmic-ray escape path length from
the Galaxy; for X$_{\rm esc} \rightarrow \infty$ (closed Galaxy) Q/W
increases by about a factor of 2; for the CRI model, Q/W depends on
the ISM composition, but is independent of the evolution since we
use the indicated O/H dependence based on the Israelian et al.
(1998) data. (b) Be production as a function of time for the CRS and
CRI models; the differences for TS, SN, and SN+CS, and the mixing
times in the CRI model result from the different Fe, and
consequently O, ISM abundances (see text).}
\end{figure}

\section{Be Production and Evolution}

The Be yield per supernova depends on several factors (e.g. Ramaty
et al. 1997): the composition of both the accelerated particles and
the ambient medium, the energy spectrum of the accelerated
particles, the energy per supernova imparted to the accelerated
particles, and the interaction model for the accelerated particles,
characterized by a path length for escape from the Galaxy, X$_{\rm
esc}$. For the CRS model at all past epochs the accelerated particle
composition is identical to the current epoch cosmic-ray source
composition. For the CRI model the composition of the accelerated
particles depends on [Fe/H], being derived from the ISM composition
at the same [Fe/H] by applying the enhancement factors that modify
the current epoch ISM to yield the current epoch cosmic-ray source,
within the Ellison et al. (1997) shock acceleration theory. For the
LECR model the accelerated particle composition at all past epochs
is also identical to the current epoch cosmic-ray source
composition, except for the variant LECR(metal), for which only C
and heavier nuclei are assumed to be accelerated. The ambient medium
composition is solar, scaled with 10$^{[{\rm Fe/H}]}$, except that
for O the scaling is given by ${\rm O/H = (O/H)}_\odot 10^{0.6{\rm
[Fe/H]}}$, which provides a good fit to the data on the dependence
of the O abundance on [Fe/H] (Israelian et al. 1998, figure~6). The
accelerated particle source energy spectra are given by an
expression appropriate for shock acceleration, $q(E) \propto
(p^{-2.2}/\beta) {\rm exp}(-E/E_0)$, where $p, c\beta$ and $E$ are
particle momentum/nucleon, velocity and energy/nucleon,
respectively, and $E_0$ is a turnover energy that we take equal to
10 GeV/nucleon for the CRS and CRI models (implying a spectrum
extending to ultrarelativistic energies) and to various
nonrelativistic values for the LECR and LECR(metal) models.

We derive Q/W (Ramaty et al. 1997), the total number of Be nuclei
produced by an accelerated particle distribution normalized to unit
cosmic-ray energy, for a given source energy spectrum and
composition, and interacting in an ambient medium of given
composition. Q/W is shown in Figure~3a as a function of the [Fe/H]
of the ambient medium, for the CRS and CRI models. For the CRS
model, it is essentially constant for [Fe/H]$< -1$, increasing
slowly thereafter, but by no more than a factor of 2. For the CRI
model, Q/W is a strong function of [Fe/H], increasing from a value
at [Fe/H]= $-3$ that is almost 2 orders of magnitude smaller than
the corresponding Q/W for the CRS model.

\begin{figure}
\plottwo{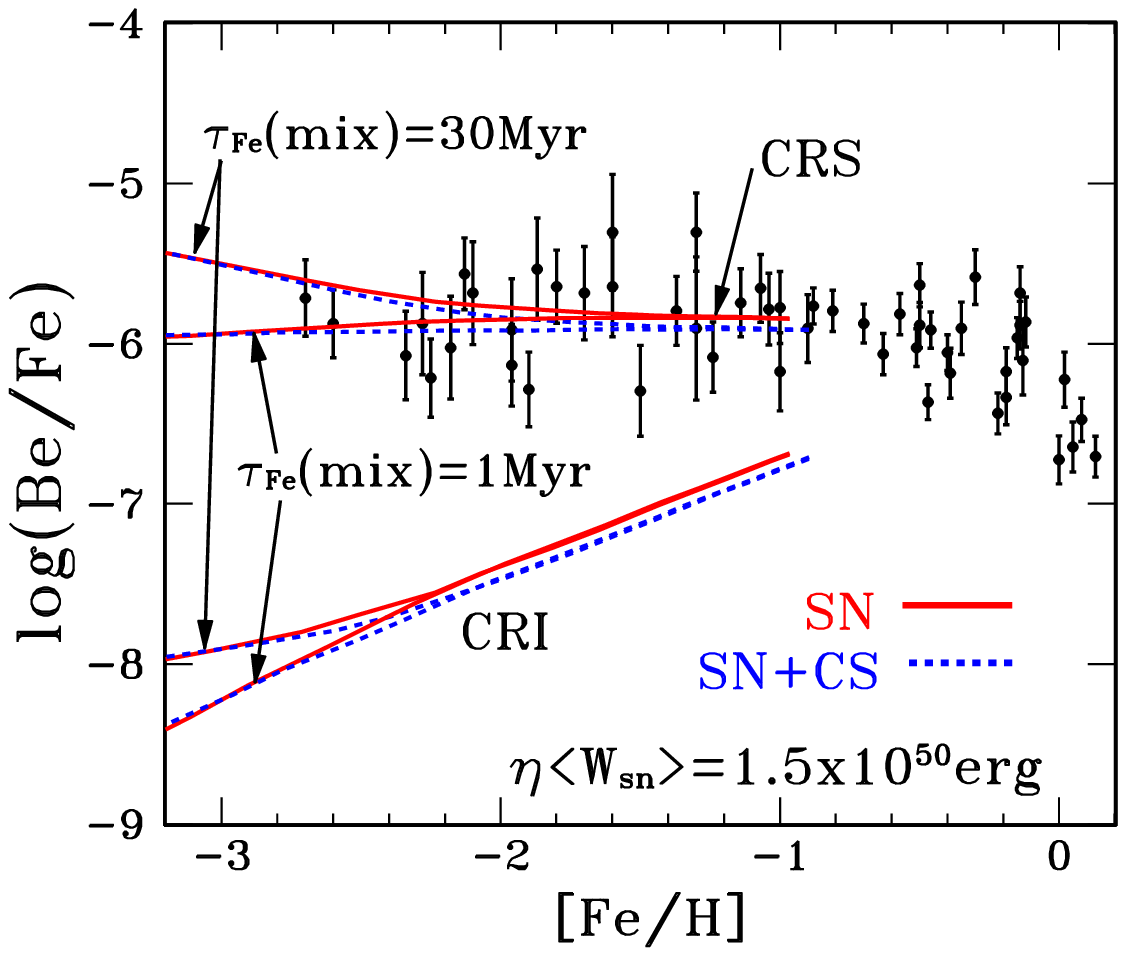}{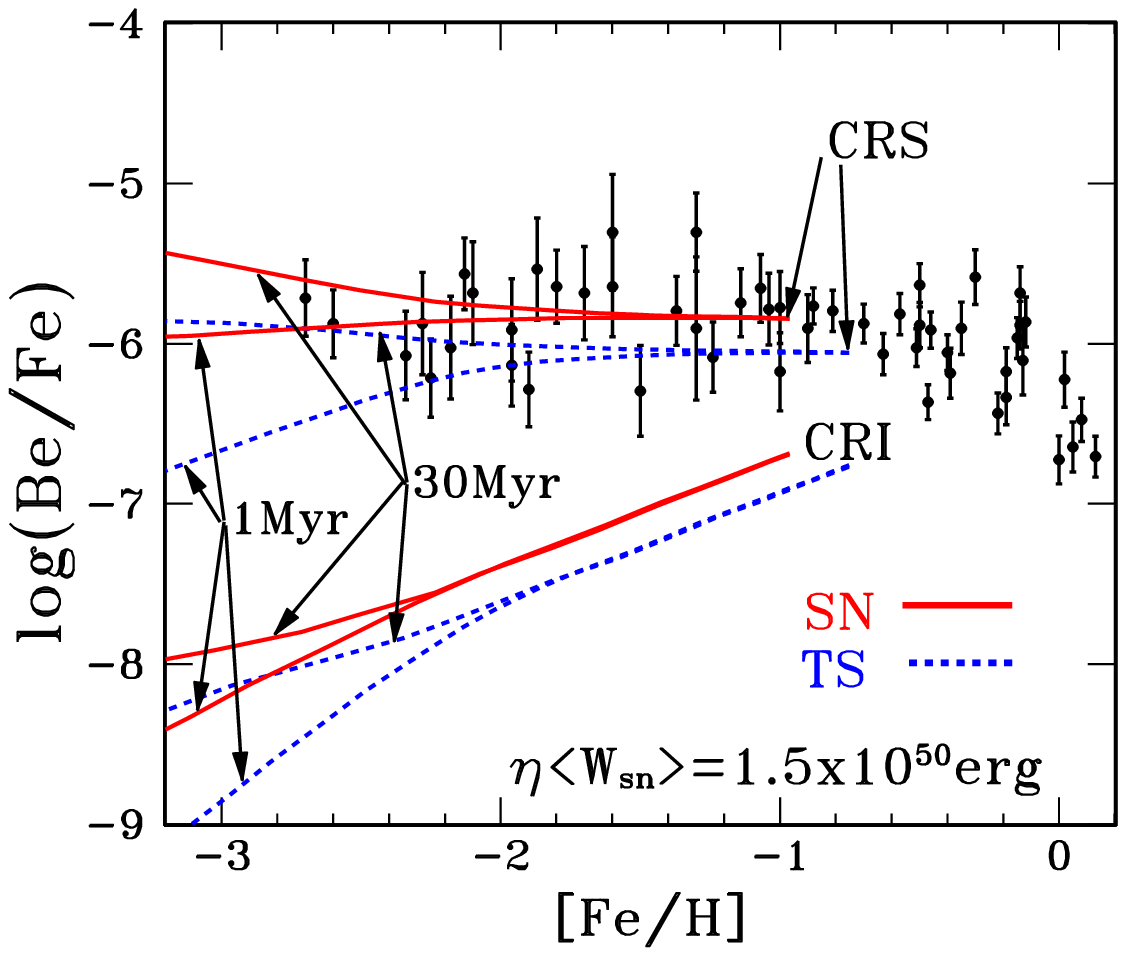}
\caption{Be/Fe evolution for 2 models, 3 cases and 2
values of the Fe mixing time. The cosmic-ray composition is
independent of [Fe/H] for the CRS model and is that of the ISM,
modified by shock acceleration theory, for the CRI model. The Fe
ejecta as a function of supernova progenitor mass are those of WW95
for the SN case, of collapsar (hypernova) augmented WW95 ejecta in
the SN+CS case, and of Shigeyama \& Tsujimoto (1998) and Tsujimoto
\& Shigeyama (1998) for the TS case. The average energy supplied to
the cosmic rays for all models and cases, 1.5$\times$10$^{50}$ erg,
is the same as that required for the current epoch cosmic rays. The
evolution is limited to 1 Gyr corresponding to [Fe/H]$\simeq -1$.
The decrease of Be/Fe at higher [Fe/H] is due to the Type Ia
supernovae which are not included. Data compilation by VF98.}
\end{figure}

The Be production per supernova is (Q/W)$\eta$W$_{\rm SN}$, where
$\eta$ is the acceleration efficiency. As already detailed, the
ejecta kinetic energy, W$_{\rm SN}, is 1.5 \times 10^{51}$ erg for
all supernovae in the TS and SN cases and for 97\% of the supernovae
in the SN+CS case.  For the remaining 3\% (the collapsars), W$_{\rm
SN} = 2.8 \times 10^{52}$ erg. Since both CRS and CRI are current
epoch cosmic-ray models, $\eta$$<$W$_{\rm SN}$$>$ must equal the
cosmic-ray energy input per supernova, $\simeq$1.5$\times 10^{50}$
erg (e.g. Lingenfelter et al. 1998). We take $\eta$ independent of
progenitor mass and the average over the employed supernova
ensemble. This implies that $\eta =0.1$ for both the TS and SN
cases, but, because of the extra energy due to the collapsars, $\eta
=0.06$ for the SN+CS case. We subsequently delay the Be deposition
into the ISM because of the finite propagation and interaction time
of the accelerated particles. The delay depends on the ISM density,
and on the energy spectrum and composition of the accelerated
particles. Using our Be production code (Ramaty et al. 1997), we
derived the appropriate distributions, which for an ISM density of 1
atom cm$^{-3}$ correspond to an average delay of several Myr for
both the CRS and CRI models. All our calculations are for such delay
times, except that for the LECR models the delay is negligible. The
resultant Be production rates as function of time are shown in
Figure~3b for the TS, SN and SN+CS cases, and for the two Fe mixing
times. We see that for the CRI model at early times, there is
significant dependence of the Be production on the employed case and
on the Fe mixing time, due to the dependence of the Be production on
the ISM composition. Specifically, since the Fe abundance at early
times is increasingly larger for the SN, SN+CS and TS cases,
respectively (Figure~1b), the corresponding O abundance is also
larger, leading to larger values of Q/W (Figure~3a), and thus to
larger Be production. For essentially the same reason, the Be
production is also larger for the shorter Fe mixing time. These
dependencies are absent for the CRS model, since here the Be
production is independent of the ambient medium composition. The
small increase for the SN+CS case at early times is due to the extra
available collapsar energy.

The evolution of Be/Fe for the CRS and CRI models, the SN, SN+CS and
TS ejecta cases, and the two Fe mixing times, is shown in Figure~4.
We see that the predictions of CRS model, for both the SN and SN+CS
cases (panels a and b), and the TS case (panel b), provide good fits
to the data, independent of the Fe mixing time. The fact that the
implied energy in cosmic rays per supernova, 1.5 $\times$ 10$^{50}$
erg, is very nearly the same as the value obtained from current
epoch cosmic-ray data, provides strong evidence for the validity of
the CRS model. On the other hand, for the CRI model the predicted
Be/Fe does not fit the data. This result differs from that
of Fields \& Olive (1999a, hereafter OF) who concluded, for their
best set of parameters, that the excess energy per supernova needed
at [Fe/H]= $-3$ only exceeds the current value by about a factor of
5 (see figure 9 in OF), whereas our calculations imply a discrepancy
of about 100. To understand the origin of the factor of 20
difference, we first note that OF modified the VF98 Be data so that
their approximation at [Fe/H=$-3$, log(Be/Fe)$\simeq$$-6.4$ (see
figure 3 in OF), is lower by a factor of 4 than the constant value
of $-5.84$ (VF98) which we use. The rest of the discrepancy is
probably caused by differences in the employed Fe ejected masses,
particularly those relevant at the lowest metallicities. These
masses are not quoted by OF, but from their equation (30) we
conclude that the ejected Fe/O decreases with decreasing [Fe/H] as
10$^{0.31{\rm [Fe/H]}}$, so that at [Fe/H]=$-3$ Fe/O is lowered by
factors of 4 and 9, relative the corresponding values at
[Fe/H]=$-1$ and $0$, respectively. But from our Figure~1, we see
that for all cases and mixing times, at times later than those
needed to achieve [Fe/H]=$-3$, the Fe/O production ratio is
essentially constant. So it appears that OF could have concluded
that the energetics do not present a problem for the CRI model
because they have modified the Be data and they chose to employ
ejected Fe masses which, even though not listed in their paper,
appear to be quite small at early times. On the other hand, we have
employed both calculated and measured Fe masses, and showed that
these are consistent with the O/Fe abundance data, particularly if
delayed mixing of Fe into the ISM is taken into account.

\begin{figure}
\plottwo{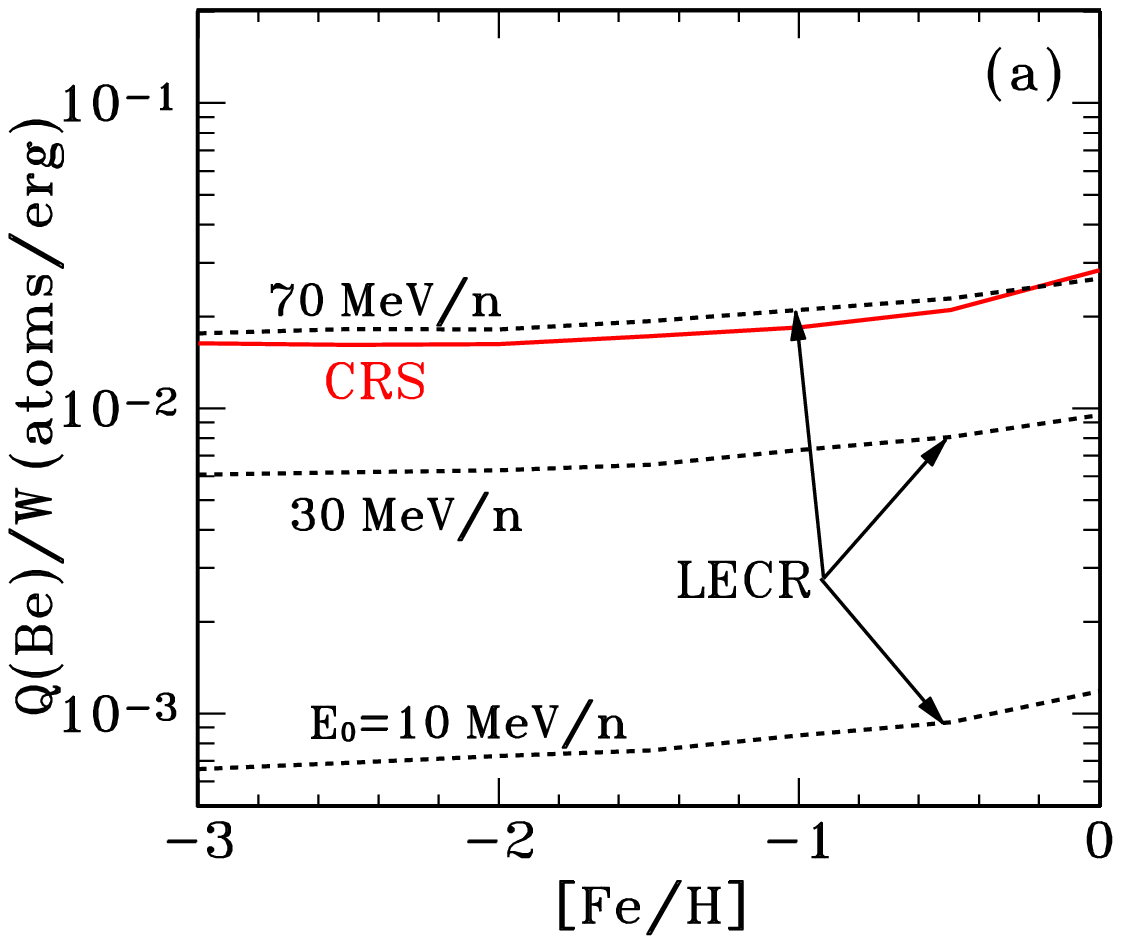}{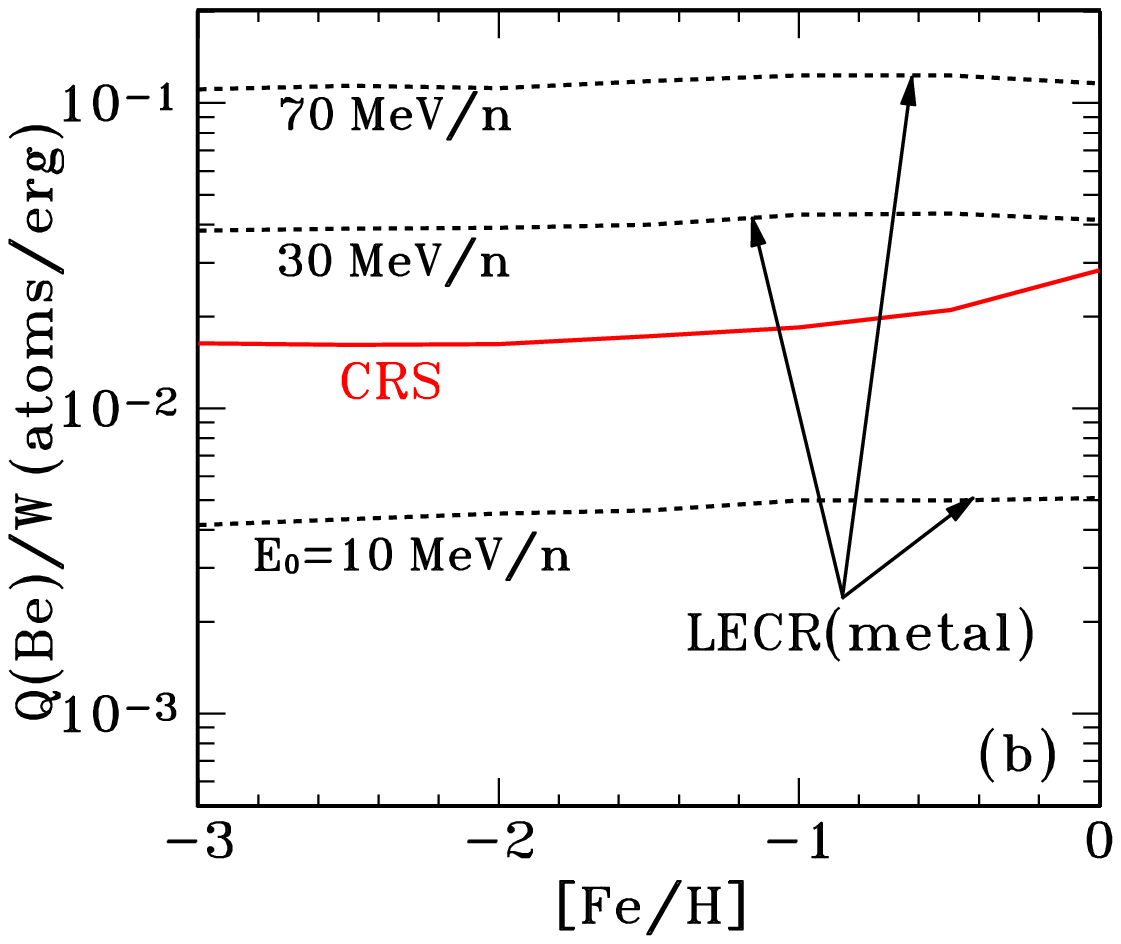}
\caption{Number of Be atoms produced per unit available
energy for the LECR and LECR(metal) models. Both the LECR and
LECR(metal) compositions are identical to that of the cosmic rays in
the CRS model, except that protons and $\alpha$ particles are absent
for LECR(metal). $E_0$ is the turnover energy that effectively cuts
off the spectrum at high energies.}
\end{figure}

We carried out similar Be production calculations for the LECR
models. The resulting Q/W, shown in Figure~5, while essentially
independent of [Fe/H], decrease rapidly with decreasing $E_0$, the
turnover energy that defines the high energy cutoff of the LECR
spectrum. At 70 MeV/nucleon, Q/W is close to its maximum (see
Ramaty et al. 1997, Figure 8). Even though the evolutionary
curves, log(Be/Fe) vs. [Fe/H] (not shown), provide good fits to
the data, the required acceleration efficiency $\eta$ becomes
prohibitively large in some cases. We have derived $\eta$ by
normalizing the evolutionary curves at [Fe/H]$\simeq -1$ to
log(Be/Fe) = $-5.84$, the best fitting constant to the Be/Fe data
for [Fe/H]$< -1$ (VF98), and by using the values of W$_{\rm SN}$
detailed above for the SN, SN+CS and TS cases, the values of Q/W
for the 3 turnover energies for the LECR and LECR(metal) models in
Figure~5, and, exploring the suggestion (VF98 and references
therein) that only supernovae from the most massive progenitors
($>$60 M$_\odot$) contribute to LECR acceleration, for 3 such
low-mass cutoffs (M$_{\rm cut}$). The results are given in
Table~2. By requiring that the efficiency $\eta$ be significantly
less than 1, we see that if the LECRs were to be the dominant
source of Be in the early Galaxy, M$_{\rm cut}$ = 60 M$_\odot$ is
only possible if the acceleration of protons and $\alpha$
particles is suppressed and if $E_0 >$ 30 MeV/nucleon. Likewise,
acceleration of metals only is also required if $E_0$ were 10
MeV/nucleon. Clearly, nuclear gamma-ray line measurements are
needed to determine the intensity of a possible LECR component and
whether it could be a significant contributor to spallogenic light
element production.

\begin{table}
\caption{Required acceleration efficiencies (fractions of
1.5$\times$10$^{51}$erg/supernova or
2.8$\times$10$^{52}$erg/collapsar) for the LECR and LECR(metal)
models for 3 values of $E_0$ (in MeV/nucleon) and 3 low mass
cutoffs for the supernova progenitors that accelerate LECRs; for
each $E_0$ and M$_{\rm cut}$, the 3 values correspond to the SN,
SN+CS and TS cases, respectively.}\label{tbl-1}
\begin{center}
\vskip -0.2truecm
\scriptsize
\begin{tabular}{cccccccccccc}
$E_0$ &M$_{\rm cut}$=10M$_\odot$
&M$_{\rm cut}$=30M$_\odot$&M$_{\rm cut}$=60$_\odot$ \\
\tableline
& & LECR: p,$\alpha$,metals\\
10 & 2.12 1.62 3.48 & 10.8 3.26 17.7 & 43.7 53.6 71.5\\
30 & 0.24 0.18 0.39 & 1.22 0.37 2.00 & 4.96 6.09 8.15 \\
70 & 0.08 0.06 0.14 & 0.42 0.13 0.69 & 1.71 2.10 2.81 \\
& & LECR(metal): metals only\\
10 & 0.35 0.27 0.58 & 1.80 0.54 2.95 & 7.31 8.99 12.0\\
30 & 0.04 0.03 0.06 & 0.19 0.06 0.32 & 0.79 0.97 1.30\\
70 &0.01 0.01 0.02 & 0.07 0.02 0.11 & 0.28 0.34 0.46 \\
\end{tabular}
\vskip -1 truecm
\end{center}
\end{table}

\section{CRI vs. CRS for the Current Epoch Cosmic Rays}

The arguments that have been made (e.g. Meyer et al. 1997) to
support cosmic-ray acceleration out of average ISM material rather
than from fresh nucleosynthetic matter (the CRI vs. CRS paradigm),
and the counter-arguments (Lingenfelter et al. 1998), based on
acceleration in supernova ejecta-enriched superbubbles (Higdon et
al. 1998) are the following:

i) {\it The enrichment of the cosmic-ray source abundances relative
to solar abundances in refractory elements relative to volatile
elements.} But the refractory grains formed in supernova ejecta are
depleted in volatiles, as are the grains in the average ISM, and
since the freshly formed grains still move at high velocities in the
superbubbles, their erosion products are much more readily
accelerated than volatile ions of the ambient gas.

ii) {\it The similarity of the cosmic-ray source and solar abundance
ratios of refractory elements, mainly Mg, Al, Si, Ca, and Fe.} But
the nucleosynthetic yields of these elements, averaged over initial
mass function and the relative contributions of the various
supernova types, are consistent with both cosmic-ray source and
solar abundances, which in fact the supernovae have produced.

iii) {\it The presence of s-elements in the cosmic rays, which are
not synthesized in supernova explosions.} But s-elements are present
in supernova ejecta along with the other much more abundant products
of pre-supernova burning since they are made in the cores of the
pre-supernova stars and are ejected both in the explosions and in
strong stellar winds. Very significant overabundances of the
prominent s-process products, Sr and Ba, relative to Fe were
observed in SN1987A (e.g. Mazzali, Lucy \& Butler 1992). In addition
to the presence of s-elements, the observed enrichment of r-process
nuclei in the cosmic rays, especially the strong Pt peak (Waddington
1996), provides direct support for a supernova origin.

iv) {\it The delayed acceleration of the cosmic rays.} This is based
on the recent ACE measurements (Binns et al. 1999) of cosmic ray
$^{59}$Co and $^{59}$Ni, showing that these isotopes could not have
been accelerated from fresh supernova ejecta less than $7.5 \times
10^4$ yrs after the explosion. This result can only argue against
cosmic-ray acceleration of supernova ejecta by the shock of the same
supernova (Lingenfelter et al. 1998), and not the collective
acceleration by successive supernova shocks of ejecta-enriched
matter in the interiors of superbubbles (Higdon et al. 1998) where
supernova explosions occur every 10$^5$ yr. These hot, low-density
bubbles, which reach dimensions of several hundred pc, are generated
by the winds and ejecta of supernova explosions of massive stars
formed in giant molecular cloud OB associations, that last for tens
of Myr. Since these bubbles expand with shell velocities much faster
than the dispersion velocities of the O and B star progenitors of
the supernovae that power them, the bulk of the supernovae occur in
their cores. The expanding remnants of each of these supernovae only
fill $<$1\% of this core before they slow down to sonic velocities.
Thus, the bulk of these supernovae remnants, together with their
metal-rich grain and gas ejecta, are well confined within the cores
of superbubbles. Consequently, the bulk of the metals of this hot
phase of the ISM originate from recent nucleosynthesis, within the
lifetime of the superbubble, and thus multiple supernova shocks
(every $\sim$ 10$^5$ yr) accelerate accumulated supernova ejecta on
average time scales at least as long as the delay implied by the
$^{59}$Co and $^{59}$Ni data.

In addition, cosmic-ray source injection from refractory grains
formed in supernova ejecta can explain the longstanding puzzle of
the cosmic-ray C and O abundances. The cosmic-ray source C/Fe and
O/Fe are much lower than both the solar values and those in the icy
grains (Savage \& Sembach 1996) that contain a major fraction of the
C and O in the ISM. O/Fe is lower than the solar value because only
a fraction of the O is trapped in refractory oxides, primarily
Al$_{2}$O$_{3}$, MgSiO$_{3}$, Fe$_{3}$O$_{4}$, and CaO. Similarly
C condenses in refractory grains of graphite and amorphous carbon,
as well as metal carbides, such as SiC, as have been detected in
presolar material in meteorites. These refractory O/Fe and the
supernova averaged C/Fe are in good agreement with the required
cosmic-ray source values (Lingenfelter et al. 1998).

\section{Conclusions}

To test the viability of the models that have been proposed for the
origin of the spallogenic light elements, we developed a new
evolutionary Monte-Carlo code which allows the investigation of
hitherto ignored effects, the delayed mixing into the interstellar
medium (ISM) of the synthesized Fe due to its incorporation into
high velocity dust grains and the relationship between the supernova
nucleosynthetic yields and the ejecta kinetic energy, including
events with very large ejecta energies (e.g. collapsars/hypernovae).
The models are: the traditional CRI, in which the cosmic rays are
accelerated out of an ISM which is increasingly metal poor at early
times; the CRS, in which cosmic rays with composition similar to
that of the current epoch cosmic rays are accelerated out of fresh
supernova ejecta and thus maintain a constant metallicity at all
epochs; and the LECR, in which a postulated low energy component,
also accelerated out of fresh ejecta, coexists with the CRI cosmic
rays.

By including the delayed mixing, we find that several sets of O
and Fe ejecta yields, based on both observations and
nucleosynthetic calculations, can account for the new data that
indicate a monotonically increasing [O/Fe] with decreasing [Fe/H].
We provide arguments showing that the bulk of the synthesized Fe
probably resides in refractory dust grains which could travel much
farther before they came to rest than the volatiles trapped in the
plasma which are more rapidly slowed by the swept up gas and
magnetic fields. Thus, there could be a delay between the
effective deposition times into the ISM of refractories, such as
Fe, and volatiles such as O, of which only a fraction is condensed
in oxide grains. We suggest that this difference in volatility
between the synthesized O and Fe is a likely cause for the
increase of [O/Fe] with decreasing [Fe/H]. In support of this
suggestion we point out that, unlike O, there appears to be no
increase with decreasing [Fe/H] of the abundances of the $\alpha$
nuclei Mg, Si, Ca and Ti relative to Fe below [Fe/H]=$-1$ (Ryan et
al. 1996), consistent with the fact that these elements are also
refractory, and thus should mix in the same way as Fe.

We use three sets of O and Fe nucleosynthetic yields, their
corresponding ejecta kinetic energies, and a cosmic-ray
acceleration efficiency consistent with current epoch cosmic-ray
data, to show that the CRI model significantly under-predicts the
measured Be abundance at the lowest [Fe/H], the increase in [O/Fe]
notwithstanding. On the other hand, the CRS model can account for
the measured Be abundances with an acceleration efficiency that is
in good agreement with the cosmic-ray data. To further compare
these models, we review the arguments against cosmic-ray
acceleration out of supernova ejecta and indicate how they are
resolved in the CRS model by averaging the nucleosynthetic yields
over initial mass function and the relative contributions of the
various supernova types, by accelerating grain erosion products in
supernova produced superbubbles, and by taking into account the
products of burning in the cores of pre-supernova stars. As
suggested by Westphal et al. (1998), the measurement of cosmic-ray
actinides with lifetimes comparable to the age of the cosmic rays
(a few tens of Myr) could distinguish between the CRI and CRS
models.

The LECR model can also account for the observations, except that the
required acceleration efficiency becomes prohibitively large for some of
the parameters that have been previously assumed for this as-yet
undetected cosmic-ray component. Future gamma-ray line observations could
establish the contribution of LECRs to the origin of the spallogenic light
elements.

\end{document}